\documentclass[sigconf,natbib=true,anonymous=false]{acmart} 

\copyrightyear{2026}
\acmYear{2026}
\setcopyright{cc}
\setcctype{by}
\acmConference[ICTIR '26]{Proceedings of the 2026 International ACM SIGIR Conference on Innovative Concepts and Theories in Information Retrieval (ICTIR)}{July 25, 2026}{Melbourne, VIC, Australia}
\acmBooktitle{Proceedings of the 2026 International ACM SIGIR Conference on Innovative Concepts and Theories in Information Retrieval (ICTIR) (ICTIR '26), July 25, 2026, Melbourne, VIC, Australia}
\acmDOI{10.1145/3805713.3820436}
\acmISBN{979-8-4007-2600-2/2026/07}
\usepackage{amsmath}
\usepackage{graphicx}
\usepackage{booktabs}
\usepackage{multirow}
\usepackage{enumitem}
\usepackage{listings}
\usepackage{tcolorbox}
\usepackage{microtype}
\usepackage{float}
\usepackage{todonotes}
\usepackage{siunitx}
\setlist[itemize]{noitemsep, topsep=0pt, leftmargin=*}
\setlist[enumerate]{noitemsep, topsep=0pt, leftmargin=*}
\newcommand{\runinheading}[1]{\noindent\textbf{#1}\quad}

\title{Adaptive Re-Ranking}

\author{Ata Cinar Genc}
\authornote{Both authors contributed equally to this research.}
\affiliation{%
  \institution{University of Massachusetts Amherst}
  \city{Amherst}
  \state{MA}
  \country{USA}
}
\email{agenc@umass.edu}

\author{Emir Kaan Korukluoglu}
\authornotemark[1]
\affiliation{%
  \institution{University of Massachusetts Amherst}
  \city{Amherst}
  \state{MA}
  \country{USA}
}
\email{ekorukluoglu@umass.edu}

\author{James Allan}
\affiliation{%
  \institution{University of Massachusetts Amherst}
  \city{Amherst}
  \state{MA}
  \country{USA}
}
\email{allan@cs.umass.edu}

\begin{document}


\begin{abstract}

Modern Information Retrieval (IR) systems typically use a ``retrieve-then-rerank'' pipeline, where a computationally expensive, pre-determined cross-encoder re-ranks the top results from a fast initial retriever. While effective, this approach often applies heavy re-ranking models regardless of query complexity, resulting in high latency and wasted computational resources on simple queries. We propose Adaptive Re-Ranking, an utility-based labeling framework for cost-aware routing and present empirical evidence (via oracle analysis and a trained baseline router) that per-query routing offers large potential gains but is non-trivial to learn from limited supervision. We train a routing classifier with 3 strategies: sparse retrieval (BM25), dense re-ranking (MiniLM-L6-v2), and heavy neural re-ranking (BGE-v2-m3). Compared to BGE our method achieves $1.15-53$x lower median latency and $1.11-5.22$x lower mean latency across all datasets we have tested, while delivering $-17.5\%$ to $+4.0\%$ nDCG@10, which is competitive in some datasets. Our findings show that routing queries based on our novel utility function offers a scalable solution for reducing computational costs and latency in a variety of IR systems.

% a framework that dynamically routes queries to the most cost-effective strategy—ranging from sparse retrieval (BM25) and dense re-ranking (MiniLM-L6-v2) to heavy neural re-ranking (BGE-v2-m3)—based on query complexity. To train our routing classifier, we introduce a novel utility function to label queries based on their retrieval effectiveness penalized by latency. 

%Our classifier demonstrates the feasibility of this approach, achieving competitive retrieval quality while maintaining low latency per query compared to a traditional heavy reranker pipeline. Oracle experiments (assuming perfect routing) demonstrate the potential for substantial improvements in both retrieval effectiveness and speed compared to hybrid retrieval approaches.

\end{abstract}   

\maketitle

\section{Introduction}
The field of Information Retrieval (IR) has gone through a paradigm shift with the adoption of dense and cross-encoder neural networks. While traditional lexical models like BM25 \cite{Robertson_Bm25} served as the cornerstone of retrieval systems for decades due to their unparalleled efficiency, lexical models often fail to capture the semantic meanings of complex queries. 
The introduction of the Transformer architecture \cite{Vaswani_2017} has also made it possible to use Transformers for ranking, using BERT to re-rank  passages~\cite{nogueira2019passage}, using sentence-embeddings for similarity checks~\cite{reimers-2019-sentence-bert}, and so on. More recently, powerful cross-encoder rerankers, such as BGE-reranker-v2-m3 (BGE) \cite{chen2024bge}, have significantly improved retrieval accuracy. However, these models also have some drawbacks, such as an increase in latency up to $15x$ (according to our experiments).

\begin{figure*}[t!]
    \centering
    % Make sure to upload your image file and name it 'pipeline_diagram.jpg'
    \includegraphics[width=0.78\textwidth]{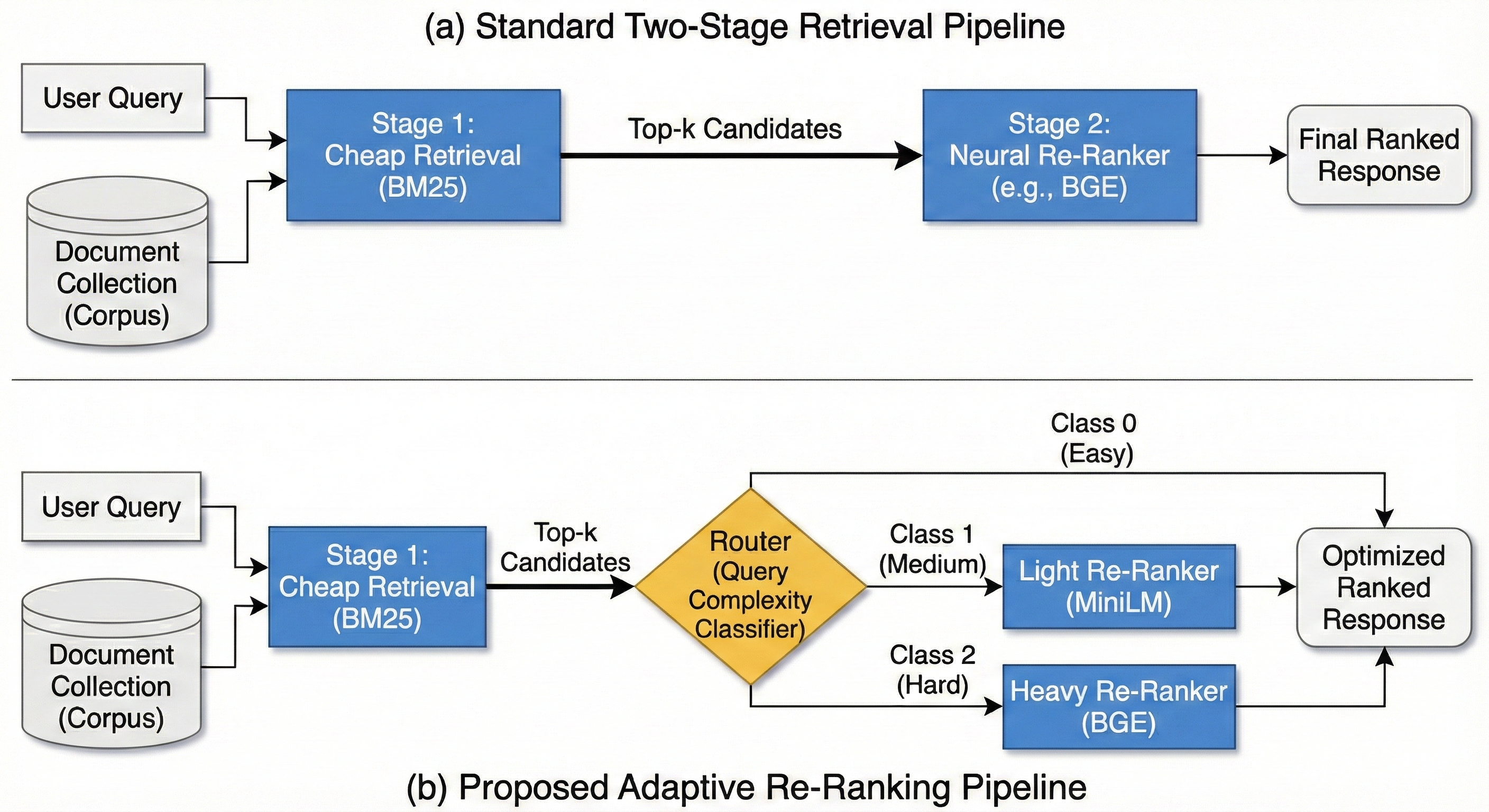} 
    \caption{System Architecture Comparison. (a) The standard two-stage  pipeline applies a heavy reranker to all queries. (b) Our  Adaptive Re-Ranking pipeline uses a lightweight Router to decide the most cost-effective path (Class 0, 1, or 2) per query.}
    \label{fig:pipeline_architecture}
\end{figure*}

Modern ad-hoc passage retrieval pipelines deployed at scale in web 
search, enterprise search, and question-answering systems are typically 
structured as a two-stage process (Figure~\ref{fig:pipeline_architecture}a): a fast, lexical retriever followed by a computationally expensive cross-encoder. We argue that using the most effective ``heavy'' cross-encoder for every single query might result in drastically high average latency, as well as enormous computational cost, since this pipeline treats every query equally. Our analysis 
of training data reveals that this uniform approach is wasteful: 
approximately 40\% of queries show no benefit from re-ranking at all, 
and only 11\% benefit from using a heavy re-ranker over using BM25 only or BM25 paired with a lightweight reranker. This shows that a significant portion of user queries is simple enough to be accurately handled by a fast, low-cost method like BM25 or a lightweight reranking model. Using a large reranking model on every query constitutes a huge waste of computational resources. However, while using a heavier re-ranker might seem unnecessary in the first place, on domain-specific datasets it provides higher nDCG@10, and it outperforms the light re-ranker. For instance, on the Arguana dataset, the heavier reranker outperforms the lighter one with a $30\%$ increase in nDCG@10.
 
To address this efficiency bottleneck, we propose \textbf{Adaptive Re-Ranking}---a cost-aware framework built on a simple yet effective principle: if the expected effectiveness gain from the expensive ranker is not justified for the given query, a cheaper alternative should be used instead. This principle can be used with any model combination; we use BM25, MiniLM-L6 (L6), and BGE in our case study to observe the difference between models that vary significantly in size. Contributions of our paper include:

\begin{enumerate}[leftmargin=*]
    \item \textbf{A novel utility-based labeling} that defines optimal strategy selection through a trade-off between retrieval effectiveness (nDCG@10, and MRR@10) and a latency penalty, enabling dataset creation according to the desired latency penalty.
        
    \item \textbf{An empirical study} that shows adaptive routing can significantly reduce latency while maintaining competitive retrieval quality, validated across diverse datasets.
    \item \textbf{An efficiency-aware routing framework} that routes each query to one of three efficiency-aware pipelines:
    \begin{itemize}[leftmargin=*]
        \item \textbf{Class 0 (No Reranker):} Fastest execution with minimal computation using a lexical model.
        \item \textbf{Class 1 (Light Reranker):} Balanced effectiveness and latency.
        \item \textbf{Class 2 (Heavy Reranker):} Highest effectiveness at increased computation and latency cost.
    \end{itemize}
    While we experiment with 3 classes, our framework can be adjusted to any number of classes.

\end{enumerate}
\begin{comment}

Our adaptive re-ranking framework offers significant practical value for resource-constrained development environments such as mobile edge computing (MEC). Recent work has ~\cite{zhang2023resource} demonstrated that mobile devices face severe constraints in CPU, GPU, memory, and battery life. In such limited-resource environment, using heavy neural re-ranking models for all queries becomes impractical and wasteful. By dynamically routing queries, our classifier enables IR systems to achieve competitive retrieval quality while respecting computational and energy limitations. By adjusting $\lambda$ in our utility function, speed and efficiency can be prioritized for such environments.
\end{comment}
Our code and data are available at \url{https://github.com/emirkaan5/adaptive-reranker}

\vspace{-0.5em}
\section{Related Work}
\begin{comment}
 \paragraph{Efficient Neural Re-Ranking}
To facilitate faster inference without sacrificing the deep semantic understanding of well-known Transformers, knowledge distillation techniques have been widely adopted. MiniLM \cite{wang2020minilm} demonstrates that distilling the self-attention distributions of deep models into smaller student models allows for significant compression while maintaining competitive performance on ranking tasks.
Early efforts to bridge the efficiency-effectiveness gap focused on architectural optimizations. \cite{hofstaetter2020interpretable} introduced the TK (Transformer-Kernel) model, which utilizes a lightweight Transformer architecture to provide fast contextualization within strict time budgets (e.g., <200ms). However, these models apply a uniform computational load to every query, regardless of whether a simpler lexical match would have sufficed. Our work builds on this by acknowledging that query complexity is heterogeneous.

\paragraph{Query Performance Prediction (QPP)}
\end{comment}

\begin{comment}
    
\paragraph{Neural Re-Ranking}
To facilitate faster inference without sacrificing the deep semantic understanding of well-known Transformers, knowledge distillation techniques have been widely adopted. MiniLM \cite{wang2020minilm} demonstrates that with distilling self-attention into smaller models, competitive performance in ranking tasks can be maintained at a smaller size.
\cite{hofstaetter2020interpretable} Proposes a model to reduce per-query cost; however, they apply the same method to every query.
\end{comment}
\paragraph{Query Performance Prediction}
QPP's goal is to estimate retrieval effectiveness without any relevance judgments. Traditional QPP methods rely on clarity-based \citep{cronen2002predicting}, robustness-based \citep{zhou2007query}, and score-based \cite{tao2014query} approaches that uses corpus statistics and query term frequencies as signals. 
Recent advances have introduced neural approaches to QPP such as  an end-to-end neural framework \cite{zamani2018neural}. The introduction of pre-trained language models further advanced the developments  by fine-tuning BERT for QPP, achieving state-of-the-art performance with significantly lower latency than previous neural predictors \cite{arabzadeh2021bertqpp}. Recent work has also explored using LLMs for QPP by, for example, fine-tuning open-source LLMs to generate relevance judgments to estimate effectiveness~\citep{meng2024qpp}. Despite these advances, prior work found that LLMs are not reliable enough to create relevance judgments \cite{faggioli2023llmjudgment}. In section \ref{sec:simple_signals}, we investigate whether post-retrieval signals from BM25 that is used in QPP (e.g., score distributions) can predict the re-ranking process for all datasets; however, we find that they are insufficient and not generalizable for per-query routing for all datasets. Our work differs from traditional QPP: rather than predicting a performance score, we are routing a classification problem over re-ranking strategies, and we include \textbf{latency} into our novel utility-based labeling function. 

\paragraph{Multi-stage and Cascade Ranking}
The retrieve-then-rerank architecture is an example of cascade
ranking, where re-rankers of increasing cost and accuracy are chained so that
expensive models only score a small candidate set produced by cheaper
stages. Foundational work framed this as a way to manage the
efficiency$-$effectiveness tradeoff \cite{wang2011cascade}. Subsequent work made the cascade explicitly cost-aware \cite{chen2017costaware}, predicted efficiency$-$effectiveness tradeoffs across stages without requiring relevance judgments \cite{clarke2016assessing}. These approaches have the cascade structure fixed, therefore going through the same sequence of stages, and optimization in this comes from pruning the candidate set, not changing the path the query takes every time. Our work differs by making the re-ranking sequence a per-query decision. Our router decision is based on the given query and the expected latency cost of using a least efficient approach.

\paragraph{Efficient Retrieval via Dynamic Pruning}
A complementary line of work reduces cost at the first-stage retrieval layer
rather than at re-ranking. Dynamic pruning algorithms such as WAND
\cite{broder2003wand} and Block-Max WAND \cite{ding2011blockmax} skip
documents that provably cannot reach the current top-$k$ threshold,
substantially reducing the number of postings scored while returning the same
results. Selective pruning extends this idea by varying pruning aggressiveness
on a per-query basis \cite{tonellotto2013selective}, which is conceptually
close to our own per-query adaptivity but applied to candidate generation
rather than re-ranking. These methods are complementary to our router as they speed up the BM25 stage, which all paths use, therefore complementing overall latencies across all paths.

\paragraph{Adaptive Retrieval}
Adaptive retrieval methods use learned signals to choose between sparse, dense, and hybrid pipelines \cite{arabzadeh2021predictingefficiencyeffectivenesstradeoffsdense}. Other work claims that the first-stage ranker is the bottleneck in latency and proposes a unified prediction framework that chooses hyperparameters based on the given query \citep{10.1145/3159652.3159676}. More recently, a new paradigm has been explored that is similar to mixture of experts in structure and aims to leverage a zero-shot mixture of heterogeneous retrievers—including human-like sources—that dynamically weights and fuses them per query, yielding substantially better retrieval performance across diverse information needs \citep{kalra2025mor}. Our work is similar in the way of offering a selection of retrievers, however we also consider the efficiency of the ideal reranker on a per-query basis.
\paragraph{Adaptive RAG Techniques}
With the recent advances in Retrieval-Augmented Generation (RAG), new adaptive pipelines were introduced. Recent work employs a classifier to determine if RAG is necessary for a given prompt 
\cite{AdaptiveRAg}. Similarly, MBA-RAG \cite{tang2025mbarag} uses reinforcement learning to optimize retrieval strategies. Other work proposes an adaptive router for different RAG paths (symbolic, neural, hybrid) \cite{hakim2025symrag}. Our work differs by focusing specifically on the re-ranking strategies in standard IR pipelines, introducing a supervised learning approach grounded in a cost-benefit (utility) function rather than relying on LLM-generated complexity signals.

\paragraph{Early Exit Cross-Encoders}
Recent work also explored reducing the cross-encoder cost by terminating the computation early for queries that can be confidently scored with fewer layers \cite{early-exit}. While these methods reduce cost within a single re-ranker, they still use a cross-encoder for every query, and they are not supported by every model. Our approach is complementary to this method; instead of deciding how many layers to go inside a fixed re-ranker, we decide if re-ranking is needed at all, and if so whether a light or heavy re-ranker should be used. Because the two work on different levels in the pipeline, they can complement each other, where a query routed to a heavy re-ranker could exit early within it as well, which we leave to future work.

\paragraph{Routing LLMs}
With more capable, bigger LLM's and smaller, cheaper, less capable LLMs' emerging, methods like RouteLLM \cite{routellm} propose efficient router models that select between stronger and weaker LLMs during inference. Related commercial systems (e.g., OpenRouter \cite{openrouter2024auto} ) apply similar routing principles in the LLM inference domain, though they focus on LLM generation.
\section{Method}
%Blah blah blah
%Can put text here to pull the Method line to the end of the previous column it fits 2 lines
%Our method has three components: utility-based labeling, router training, and inference time routing
Our framework consists of a utility-based labeling that assigns each query to its optimal retrieval strategy, dataset curation, and router training, a routing pipeline that directs queries at inference time.

\subsection{Labeling}
All the data used for labeling is computed on the official BEIR training splits~\citep{thakur2021beir} of each dataset. Test queries are held out entirely and used for only final evaluation, and we also reserve several domain-specific datasets for only final evaluation to see how the model generalizes with zero-shot evaluation.
To create our classifier, we label queries based on what retrieval method provides the best performance with a reasonable latency. For this, we define the effectiveness score as the average of nDCG@10 and MRR@10:

    \begin{equation}
\label{eq:eff}
    \mathrm{eff}_m(q) = \frac{1}{2}\,\mathrm{nDCG@10}_m(q) + \frac{1}{2}\,\mathrm{MRR@10}_m(q).
\end{equation}
This gives a value between $(0,1)$ that captures how good the ranking is for that query, ignoring latency. The nDCG provides a graded, position‑sensitive measure of overall ranking quality for each query. However, users in modern information retrieval systems often care most about how quickly they see the first relevant result, as emphasized by prior work \cite{x}. Therefore, we additionally use MRR, which focuses on the rank of the first relevant item. Having these two metrics helps the pipeline to be able to fit into most setups since we are unaware of the query intent at labeling time; we weight both metrics equally since they represent different intents, while nDCG reflects a recall-oriented user profile who inspects the full result, and MRR reflects a precision-oriented user profile that wants a single relevant result as quickly as possible \cite{x}. Thus, using both nDCG and MRR lets us capture both holistic list quality and the user’s experience of finding the first useful result. \footnote{We have also experimented with only using nDCG@10 in labelling, the trained router provided better results with the mean of nDCG@10 and MRR@10.}

However, we do not want to choose models only by effectiveness; slower models should be penalized. While recent works  \cite{AdaptiveRAg,tang2025mbarag,arabzadeh2021predictingefficiencyeffectivenesstradeoffsdense} explore the trade-off between retrieval accuracy and computational cost, they do not include measured true latency (clock time) in decision making. To be able to train our classifier with latency included we introduce a formula that penalizes the latency \textit{per query} instead of taking the average of all queries. We are aiming to reward queries that only use more computational resources when doing so will actually improve that query itself.

Let ${lat}_m(q)$ be the latency of a given query of model $m$, and let $\max_j(\mathrm{lat}_j(q))$ be the maximum latency among all models for that query. We mix the effectiveness and latency scores using $\lambda$ to define our utility function:
\begin{equation}
\label{eq:util}
    \mathrm{util}_m(q) = (1-\lambda)\,\mathrm{eff}_m(q)
    + \lambda\left(1 - \frac{\mathrm{lat}_m(q)}{\max_j (\mathrm{lat}_j(q))}\right).
\end{equation}
Note that the model with the lowest latency will have a boosted utility score compared to the others; indeed, the most expensive model will receive no contribution to the utility from latency. Intuitively, $\mathrm{util}_m(q)$ is high when the model is both effective and fast, and it decreases when the model is slow relative to the others. 

\begin{comment}
    
The parameter $\lambda$ serves as a hyperparameter controlling the system's sensitivity to latency. After conducting sensitivity analysis, we select $\lambda = 0.05$ to enforce a `performance-first' policy\footnote{We performed sensitivity analysis with the oracle across $\lambda \in \{0.01, 0.05, 0.1, 0.15, 0.2\}$. $\lambda = 0.05$ yielded the best balance, providing the least effectiveness decline with moderate latency reductions.}. This value ensures that while slower ranking methods (such as using BM25+BGE-v2-m3) incur a penalty, they are not systematically disqualified. Consequently, this formula implies that a significant increase in latency is acceptable only if it yields a measurable gain in retrieval effectiveness, effectively acting as a regularizer against inefficient models that offer negligible performance improvements. $\lambda$ can be adjusted according to need: if the user desires the system to run faster, then it can be adjusted by increasing $\lambda$.
\end{comment}

\subsection{Hyperparameter Tuning}
\begin{figure}[H]
  \centering
  \includegraphics[width=\linewidth]{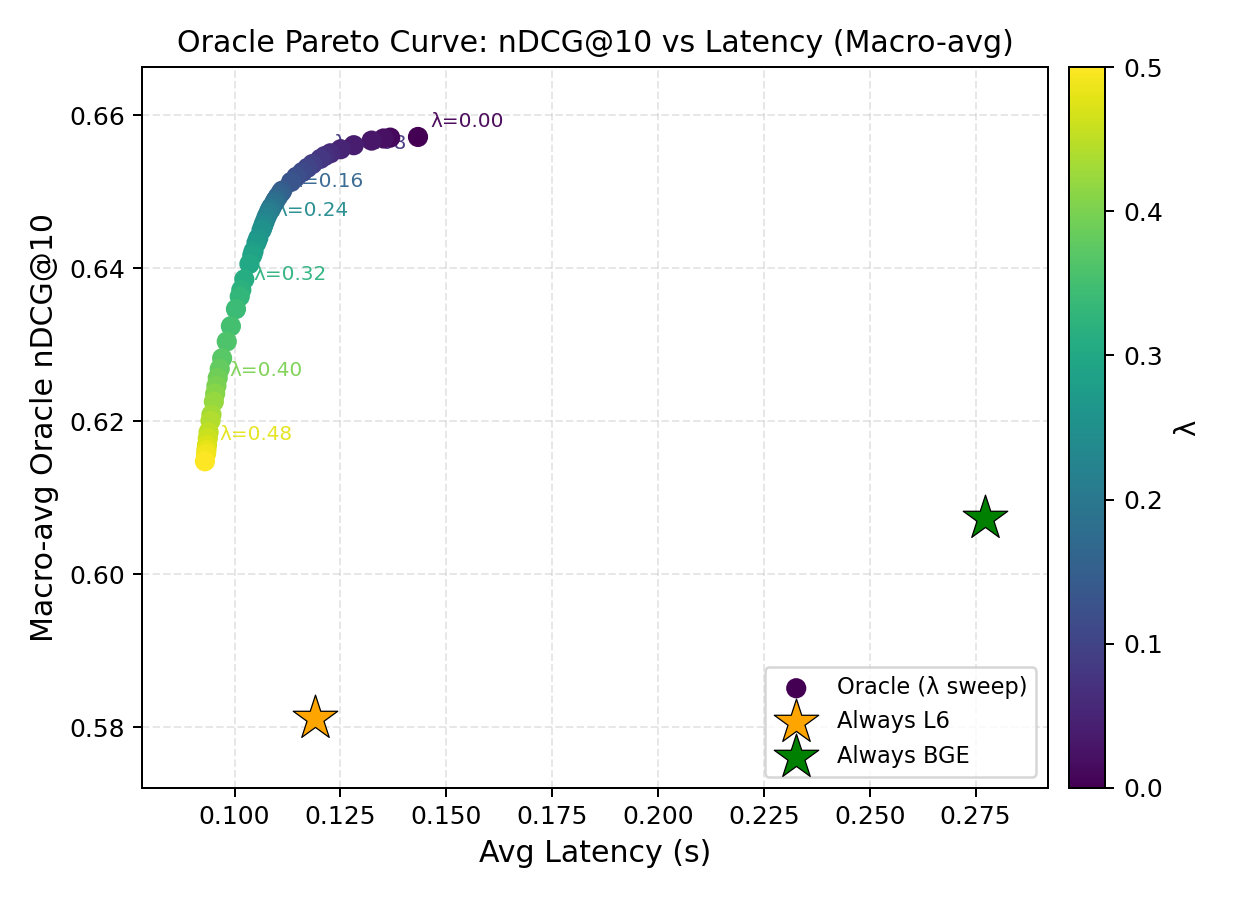}
  \caption{Oracle Pareto curve across $\lambda \in [0.0, 0.5]$ on BEIR train datasets. Each point represents a different $\lambda$ value. BM25 has an avg nDCG@10 of 0.4845 and an avg latency of 82.4ms. The oracle dominates all fixed baselines at every operating point.}
  \label{fig:oracle_pareto}
\end{figure}

The parameter $\lambda$ in our utility function (Eq.\ref{eq:util}) controls the trade-off between effectiveness and efficiency (latency). At $\lambda = 0$ the labeling selects the model with the highest effectiveness regardless of the cost. As $\lambda$ values increase, slower models are increasingly penalized; thus, the classifier would learn a more balanced approach that is cost-aware. Since this hyperparameter is applied to the dataset during pre-training step, we perform a sensitivity analysis using an oracle router (which always selects the optimal path) to find the optimal value for $\lambda$ satisfying our intent. The oracle Pareto curve (Figure \ref{fig:oracle_pareto}) shows the effectiveness-efficiency trade-off of different $\lambda$ values using Oracle on training splits of BEIR datasets. It is observable that $\lambda=0.05$ provides the least efficiency lost, with the most effectiveness when compared to other $\lambda$ values. On Oracle, $\lambda=0.05$ provides nDCG@10 of $\approx 0.655$ while having approximately 0.12s average latency. It provides higher effectiveness than always picking the most expensive model, while being less than half of its latency. As a result, we picked $\lambda=0.05$ for our experiments. With higher lambda values, we observe that class sizes skew towards lighter models, which results in insufficient data distribution for training.

\subsection{Dataset Curation}
We curated a dataset using BEIR datasets with 306,544 labeled queries with $\lambda = 0.05$. It shows a skew toward Class 0 (no reranking) of $59.2\%$ (see Table \ref{tab:label-stats}) which indicates that the majority of queries can be handled efficiently with only BM25 without increasing re-ranking latency. However, queries labeled as Class 1 (MiniLM-L6) received higher average effectiveness, even outperforming the heavy reranking in that subset. It can also be observed that in Class 2 (BGE) the average effectiveness scores of both BM25, and MiniLM-L6 is \textit{significantly} lower than BGE-v2-m3, which illustrates that for some queries, a bigger model yields more effectiveness -- but not always, which justifies our approach. Illustrating these patterns at the query level, Table \ref{tab:metrics} shows examples from each class and the utility scores across different retrieval models.
\begin{table}[h]
\centering

\small
\setlength{\tabcolsep}{9pt}
\renewcommand{\arraystretch}{1.2}
\begin{tabular}{l r c c c}

\toprule
\textbf{Label Category} & \textbf{Count} & $\mathrm{\overline{Eff}}_{\text{BM25}}$ & $\mathrm{\overline{Eff}}_{\text{L6}}$ & $\mathrm{\overline{Eff}}_{\text{BGE}}$ \\
% \cmidrule(lr){1-2} \cmidrule(lr){3-5}
%  & &  \\
\midrule
Class 0 (BM25) & 182,876 & 0.2203 & 0.1801 & 0.1831 \\
Class 1 (L6)   & 88,295 & 0.2049 & 0.7475 & 0.6467 \\
Class 2 (BGE)  & 35,573  & 0.1614 & 0.3492 & 0.6441 \\
\bottomrule
\end{tabular}

\caption{Label distribution and average effectiveness scores across the dataset, $\overline{Eff}$ denotes mean, calculated with Eq.~\eqref{eq:eff}\protect\footnotemark }  
\label{tab:label-stats}
\end{table}
\footnotetext{L6 means BM25 + L6, BGE means BM25+BGE}

\begin{comment}
    
\begin{table}[h]
\centering
\small
\setlength{\tabcolsep}{6pt}
\renewcommand{\arraystretch}{1.2}
\begin{tabular}{l r c c c}
\toprule
\textbf{Label Category} & \textbf{Count} & $\overline{\textbf{nDCG}}_{\textbf{BM25}}$ & $\overline{\textbf{nDCG}}_{\textbf{L6}}$ & $\overline{\textbf{nDCG}}_{\textbf{Gemma}}$ \\
\midrule
Class 0 (BM25)  & 88,876  & 0.2009 & 0.1652 & 0.1687 \\
Class 1 (L6)    & 45,526  & 0.2355 & 0.7470 & 0.6541 \\
Class 2 (Gemma) & 15,735  & 0.2116 & 0.4199 & 0.7207 \\
\bottomrule
\end{tabular}
\caption{Label distribution and average nDCG@10 scores per class ($\lambda = 0.08$).}
\label{tab:label-stats}
\end{table}
\end{comment}

\subsection{Retriever Decision}
We choose \texttt{BM25} as the first-stage ranker based on its proven success and popularity. For light and heavy rerankers, we use \texttt{ms-marco\-MiniLM-L6-v2} \cite{wang2020minilm} and \texttt{BAAI/bge-reranker-v2-m3}\cite{chen2024bge} , again, based on their performance and popularity \cite{huggingface_minilm_reranker, huggingface_bge_reranker}. We also explore other popular options such as \texttt{Qwen3-ReRanker-8B}, \texttt{Qwen3-ReRanker-0.6b} \cite{qwen3reranker, qwen3reranker8b} for the heavy reranker, and find that \texttt{bge-reranker-v2-m3} provides better accuracy on all the datasets we have used compared to the other models on the training data.
\begin{table}[t]
\centering
\small
\setlength{\tabcolsep}{4pt}
\begin{tabular}{l p{3.2cm} l r r r}
\toprule
         &      &     Best       & \multicolumn{3}{c}{Utility Score} \\
Dataset & Query & Model & BM25 & L6 & BGE\\
\midrule
MS-Marco & is levaquin an antibiotic & BGE & 0.04 & 0.26 & 0.40 \\
MS-Marco & what causes itching during exercise & L6 & 0.01 & 0.54 & 0.53 \\
NF Corpus & What is ‘Meat Glue’? & BM25 & 0.70 & 0.34 & 0.30 \\
NF Corpus & The Top Three DNA Protecting Spices & BM25 & 0.63 & 0.62 & 0.58 \\
Quora & How can I rule the world? & L6 & 0.29 & 0.98 & 0.95 \\

\bottomrule
\end{tabular}
\caption{Example scores for BM25, MiniLM-L6, and BGE across MS-Marco, NFCorpus, and Quora queries ($\lambda = 0.05$)}
\label{tab:metrics}
\end{table}

\subsection{Router Training}
% \subsubsection{Model overview}
We train a supervised \textbf{query router} that predicts a routing class, given a query. Each class corresponds to a retrieval route  (e.g., BM25 only, BM25 + Light Reranker, BM25 + Heavy Reranker). The routing goal is to select the most efficient route while maintaining ranking quality. Since our training data has imbalanced classes, we down-sample our dataset to match the minority class\footnote{We also experimented with class-weighted loss function instead of downsampling; however, it yielded lower routing accuracy than down-sampling.}. We use a 75/25 train-test ratio for training. We truncate queries and use fixed-length padding. Using this data, we fine tune \texttt{bert-base-uncased} with a sequence classification head \texttt{(num-labels=3)} using \texttt{AdamW} optimizer ($\eta = 3 x 10^{-5}$). We train the model for 20 epochs, and we select the best model based on validation accuracy. Final performance is reported as test accuracy of $65\%$ with [0.649, 0.660] 95\% confidence interval. This accuracy reflects the class distribution in Table~\ref {tab:label-stats}. We discuss this further in limitations.
The router adds a \textbf{overhead latency of $\approx 3.6$ ms}. While this creates a minor baseline cost, absorbing this small cost is highly preferable compared to the latency cost caused by routing every query through a fixed re-ranker.

\subsection{Experiment Setup}

\runinheading{First Stage } We implement BM25 with $k_1 = 1.2$ 
and $b = 0.75$ (default parameters, not tuned) in Python with an in-memory inverted index using Porter/Snowball stemming and NLTK tokenization. Documents are indexed as \textit{title} and \textit{text} (when available).

\runinheading{Candidate Set Size} Throughout all experiments, we rerank the top $k = 50$ documents returned by the BM25 first-stage retriever.

\runinheading{Training and Testing} We train the router on each dataset's official \textit{train} split. All labelling is computed using the train split only, and test queries were only used on test experiments.

\runinheading{ReRankers}  We use the cross-encoder reranker \textit{ms-marco-MiniLM-L-6-v2} (Sentence-Transformers \texttt{CrossEncoder}) to score each $(q,d)$ pair in the top-$50$ candidate set and then sorted documents by the predicted relevance score to generate the ranked lists.
We use \texttt{BAAI/bge-v2-m3} via FlagEmbedding (FP16 on GPU when available) to score the same $(q,d)$ pairs and then sort documents by the reranker score.

\runinheading{Hardware and Latency Measurement} All experiments were run on a single node with 8 CPU cores, 50 GB RAM, and with NVIDIA L40S GPU (48GB). Rerankers (L6 and BGE) were ran on the L40s in FP16 with batch size k = 50 (full reranking candidate list as a single forward), BM25 ran on CPU; the BERT router was run on GPU. Latencies are wallclocked, measured as per query in single-query mode. Identical hardware was used across every labelling and experimentation. 

\begin{table*}[!t]
\centering
\small
\setlength{\tabcolsep}{4pt}
\begin{tabular}{ll rrrrrr @{\hskip 18pt} ll rrrrrr}
\toprule
\multicolumn{8}{c}{\textbf{(a) In-domain}} & \multicolumn{8}{c}{\textbf{(b) Out-of-domain (except Quora)}} \\
\cmidrule(lr){1-8} \cmidrule(lr){9-16}
\textbf{Dataset} & \textbf{Metric} & \textbf{BM25} & \textbf{L6} & \textbf{BGE} & \textbf{Router} & \textbf{OH} & \textbf{Oracle} &
\textbf{Dataset} & \textbf{Metric} & \textbf{BM25} & \textbf{L6} & \textbf{BGE} & \textbf{Router} & \textbf{OH} & \textbf{Oracle} \\
\midrule
\multirow{3}{*}{FiQA}
  & nDCG@10           & .238 & .327 & .371 & .306 &     & .403
  & \multirow{3}{*}{FEVER}
  & nDCG@10           & .476 & .724 & .759 & .656 &     & .776 \\
  & $\overline{Lat}$  & 19.9 & 71.2 & 299  & 129  & 3.9 & 100
  & & $\overline{Lat}$  & 920  & 958  & 1154 & 999  & 3.9 & 967  \\
  & $\widetilde{Lat}$ & 18.2 & 70.1 & 299  & 69.1 & 3.6 & 35.4
  & & $\widetilde{Lat}$ & 751  & 787  & 980  & 820  & 3.7 & 791  \\
\cmidrule(lr){1-8} \cmidrule(lr){9-16}
\multirow{3}{*}{NFCorpus}
  & nDCG@10           & .314 & .330 & .324 & .337 &     & .370
  & \multirow{3}{*}{ArguAna}
  & nDCG@10           & .368 & .322 & .421 & .363 &     & .533 \\
  & $\overline{Lat}$  & 0.5  & 42.1 & 213  & 60.6 & 3.8 & 58.2
  & & $\overline{Lat}$  & 29.5 & 87.9 & 320  & 61.3 & 5.0 & 143  \\
  & $\widetilde{Lat}$ & 0.2  & 54.7 & 287  & 5.4  & 3.4 & 1.0
  & & $\widetilde{Lat}$ & 27.3 & 83.7 & 315  & 36.2 & 4.2 & 70.3 \\
\cmidrule(lr){1-8} \cmidrule(lr){9-16}
\multirow{3}{*}{MS MARCO}
  & nDCG@10           & .220 & .388 & .390 & .363 &     & .439
  & \multirow{3}{*}{DBPedia}
  & nDCG@10           & .279 & .403 & .407 & .357 &     & .444 \\
  & $\overline{Lat}$  & 598  & 616  & 694  & 626  & 3.5 & 614
  & & $\overline{Lat}$  & 368  & 384  & 458  & 399  & 4.3 & 409  \\
  & $\widetilde{Lat}$ & 447  & 464  & 542  & 470  & 3.4 & 459
  & & $\widetilde{Lat}$ & 241  & 257  & 329  & 286  & 3.6 & 278  \\
\cmidrule(lr){1-8} \cmidrule(lr){9-16}
\multirow{3}{*}{SciFact}
  & nDCG@10           & .658 & .676 & .720 & .697 &     & .756
  & \multirow{3}{*}{Quora}
  & nDCG@10           & .774 & .830 & .879 & .856 &     & .916 \\
  & $\overline{Lat}$  & 2.1  & 56.9 & 298  & 98.2 & 7.6 & 40.4
  & & $\overline{Lat}$  & 17.1 & 25.1 & 45.2 & 30.2 & 3.3 & 22.4 \\
  & $\widetilde{Lat}$ & 2.0  & 56.1 & 285  & 8.5  & 2.9 & 2.6
  & & $\widetilde{Lat}$ & 11.9 & 20.0 & 40.5 & 26.4 & 3.2 & 17.1 \\
\bottomrule
\end{tabular}
\caption{Performance across test splits of training datasets and out-of-domain datasets ($\lambda = 0.05$). $\overline{Lat}$ denotes mean, $\widetilde{Lat}$ denotes median latency. The \textbf{OH} column reports router decision overhead following the same convention: mean on the $\overline{Lat}$ row, median on the $\widetilde{Lat}$ row. All times in ms.}
\label{tab:results}
\end{table*}

\begin{table}[!t]
\centering
\small
\setlength{\tabcolsep}{3pt}
\begin{tabular}{@{}l cc cc cc@{}}
\toprule
& \multicolumn{2}{c}{\textbf{nDCG@10}} & \multicolumn{2}{c}{$\overline{Lat}$} & \multicolumn{2}{c}{$\widetilde{Lat}$} \\
\cmidrule(lr){2-3} \cmidrule(lr){4-5} \cmidrule(lr){6-7}
\textbf{Dataset} & \textbf{Router} & \textbf{Random} & \textbf{Router} & \textbf{Random} & \textbf{Router} & \textbf{Random} \\
\midrule
\multicolumn{7}{l}{\textit{In-domain}} \\
FiQA      & .306 & \textbf{.313 }& \textbf{129}  & 131  & \textbf{69.1} & 70.2 \\
NFCorpus  & \textbf{.337} & .322 & \textbf{60.6} & 86.1 & \textbf{5.4}  & 43.5 \\
MS MARCO  & \textbf{.363} & .333 & \textbf{626}  & 636  & \textbf{470}  & 486  \\
SciFact   & \textbf{.697} & .684 & \textbf{98.2} & 121  & \textbf{8.5}  & 56.2 \\
\cmidrule(lr){1-7}
\multicolumn{7}{l}{\textit{Out-of-domain}} \\
FEVER     & \textbf{.656} & .653 & \textbf{999}  & 1011 & \textbf{820}  & 839  \\
ArguAna   & .363 & \textbf{.370} & \textbf{61.3} & 146  & \textbf{36.2} & 84.0 \\
DBPedia   & .357 & \textbf{.363} & \textbf{399}  & 404  & 286  & \textbf{279}  \\
Quora     & \textbf{.856} & .828 & 30.2 & \textbf{29.2} & 26.4 & \textbf{26.3} \\
\bottomrule
\end{tabular}
\caption{Router vs.\ a uniform random baseline that selects each pipeline with probability $1/3$. $\overline{Lat}$ and $\widetilde{Lat}$ denote mean and median latency. All times in ms.}
\label{tab:router-vs-uniform-random}
\end{table}

\section{Results}
To measure the performance of our router against different methods, we conduct test runs in various IR datasets. For each query, we run the retrieval pipelines for each proposed method, and also run the pipeline for the best pipeline for each of the queries, the Oracle. We calculate the effectiveness (Eff) and latency scores per query.  Table \ref{tab:results} evaluate the central claim of the paper: \textbf{Selective re-ranking is valuable because queries differ in how much they benefit from costly re-ranking}.

\begin{figure}
    \centering
    \includegraphics[width=0.95\linewidth]{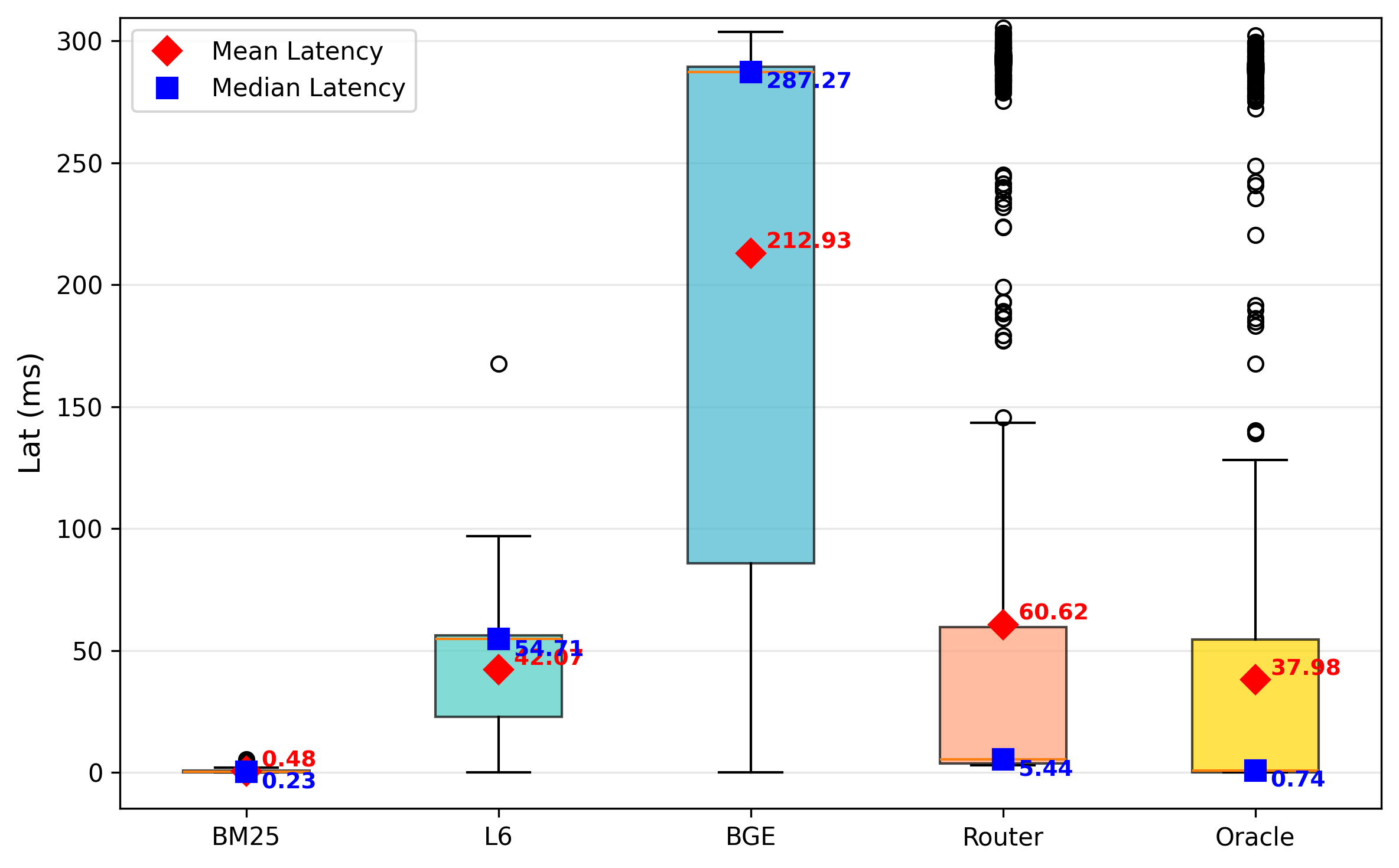}
    \caption{Latency distribution on NFCorpus}
    \label{fig:placeholder}
\end{figure}

\paragraph{\textbf{Per-Query Selection Dominates Any Fixed Strategy.}}
Table \ref{tab:label-stats} effectively portrays us how some queries benefit from certain re-rankers whilst hurting from others. Across all datasets, we see that fixed strategies always expose a latency-effectiveness tradeoff, whereas conditionally routing offers a balance. The oracle shows the strongest performance in both effectiveness and latency, confirming that adaptive routing can dominate a fixed strategy (Table \ref{tab:results}), establishing an upper bound that is unattainable by any fixed method -- e.g., on \textbf{NFCorpus} the Oracle shows 13\% higher effectiveness against both \textbf{L6} and \textbf{BGE} whilst being 5.60× faster than BGE. Since the routing includes costly methods as well, we can observe that the difference between the mean and median latency shows us that the distribution of the latency is on the lower end, which can also be seen by the figure \ref{fig:placeholder}. This shows that not only within each dataset, some queries benefit from costly routing and some do not, but also the per-query method lowers the latency distribution of query latency. 

\paragraph{\textbf{The learned router shows the applied validity of  per-query routing}}
Across all datasets, Looking at Table \ref{tab:results}, the learned router achieved 1.15×–53.2× lower median latency and 1.11×–5.22× lower mean latency, with nDCG@10 changes ranging from $-$17.5\% to $+$4.0\%. The largest wins occurred on NFCorpus and SciFact, where median latency dropped by 53.2× and 33.6× (to 5.4 ms and 8.5 ms, respectively), while effectiveness slightly increased on NFCorpus ($+$4\%) and decreased modestly on SciFact ($-$3.2\%). In Quora, the router maintained almost identical effectiveness ($-$2.6\%) while reducing the median latency by $35\%$. At the same time, the router delivers median-latency gains on \textbf{NFCorpus} (10.1×) and \textbf{SciFact} (6.60×): the router lowers the latency distribution by routing to fast lexical paths. Additionally, to investigate the gap between median and mean latency, we look at figure \ref{fig:placeholder}\footnotemark[5]. We observe that the learned router has a latency distribution very similar to the oracle. To evaluate the pipeline's robustness in real-world scenarios, we tested the router's zero-shot generalization on unseen, out-of-domain datasets (Table \ref{tab:results}b). The pipeline successfully maintained its latency advantages without requiring any domain-specific fine-tuning or new training. On Arguana, which our router was not trained on, we saw +12\% nDCG boost while achieving 26.6ms faster mean latency than L6
\footnotetext{Since all datasets had similar results, we report only NFCorpus}.These results show us that the learned router can successfully route queries effectively and cut costs where it is not needed.

% our learned router, compared to BGE, achieves \textbf{$\sim$1.15-53$x$} lower median latency across test results whilst delivering \textbf{$\sim$0.81-1.04\%} relative effectiveness. Mean latency improvements range from \textbf{$\sim$1.10-3.5$x$}. The gap between mean and median reflects the handling of most queries through fast lexical retrieval, also shown in (ref fig here). Compared to \textbf{L6}, our learned router \textbf{$\sim$1.38-195$x$} higher effectiveness  with modest latency overhead: median latency increases by 30-580\% over I6 but remains \textbf{86-98\% }faster than BGE. On \textbf{NFCorpus} and \textbf{Quora} datasets, we achieve 5\% and 38\% higher effectiveness with 98\% and 32\% faster median latency (ref fig here), respectively. 

% For example, Table \ref{tab:results} shows that in NFCorpus dataset, our trained router can be $\sim5\%$ more effective than the \textbf{BGE} whilst being 350\% faster. In the same setting our trained router is $\sim4.2\%$ more effective than \textbf{L6} whilst being 30\% slower.\todo{add median latency stuff here if all boxplots make sense. If not dont do that}

\paragraph{\textbf{Comparison against Random Baseline}}
We compare our learned router to a random routing baseline to ensure that learning query-level signals for routing queries is meaningful. We uniformly assign our test dataset to each of the 3 pipelines (BM25, BM25+L6, BM25+BGE) with 1/3 probability. From the table \ref{tab:router-vs-uniform-random}, we observe that the learned router achieves higher nDCG@10 $5/8$ times and lower mean latency $7/8$ times, substantially reducing mean and median latency. This shows us that per-query routing is a viable and learnable strategy.

\paragraph{\textbf{Addressing the Router-Oracle gap}}
The consistent gap between our learned router and the oracle indicates substantial unrealized headroom. We attribute this primarily to (i) class imbalance in the training data, with Class 2 accounting for only 11.6\% of labeled queries, and (ii) inherent label noise near the Class 1 / Class 2 boundary, where L6 and BGE utility scores are often close \ref{tab:metrics}. The router therefore tends to under-select Class 2, which explains the effectiveness drops on datasets where BGE's gains are largest (e.g., FEVER, Arguana). The oracle results demonstrate that this is a limitation of the router instance, not of the adaptive framework itself. 

\subsection{Ablation Study}

\subsubsection{Can Simple Signals Replace a Learned Router?}
\label{sec:simple_signals}

\paragraph{\textbf{Routing decision with lexical complexity.}}
To test whether lexical complexity can serve as a routing signal for the given query, we checked the queries against the Academic Word List \cite{coxhead2000} and CEFR-J advanced vocabulary (C1-C2) \cite{olp_cefrj}. We observe that for all queries on our dataset, the percentage of advanced vocabulary remained consistent, showing that  ``advanced vocabulary'' is not a valid selection criteria (AWL's Cramer's $V= 0.03$, CEFR-J C1's $V= 0.04$, and C2's $V= 0.03$). We also observe that there are no 
significant differences across classes in query length (46.3-47.2 characters, 
$p = 0.30$) and word length (5.53-5.57 characters, $p = 0.41$), so length does not play a meaningful part in the routing decision.

\paragraph{\textbf{Routing with signals from BM25}} 

We have also tested using simple signals from BM25 to try to predict which queries benefit from re-ranking, potentially not needing a router altogether. To investigate this further, we compute the Spearman rank correlations between several features derived from BM25 and compare them when we purely select re-ranking or no re-ranking according to effectiveness with the datasets we have used. We observe that all correlations are weak: The strongest signal from BM25 is the number of documents retrieved (\(\rho = 0.14\)) followed by the length of the query (\(\rho \approx 0.08\)), followed by the score gap between the top ranked and the second top ranked document (\(\rho = -0.06\)). The score gap is consistent with the intuition of "queries where BM25 finds a clear winner are less likely to benefit from a re-ranker". However, this pattern is domain-dependent. On the SciFact-train dataset, the score gap is a strong signal (9.98 for BM25 only vs 2.97 for reranked queries). However, this effect does not generalize across the other datasets, for instance, on FiQA-train, the signal was 2.10 for BM25 and 1.91 for re-ranked queries. This implies that routing decision varies on corpus, emphasizing the need for a learned router to capture query-level signals beyond BM25 statistics. Thus, we come to the conclusion that none of these signals are strong enough to build a reliable routing mechanism.

\vspace{-1ex}
\section{Conclusion}
In this paper, we propose an \textbf{Adaptive Re-Ranking} framework designed to mitigate the computational bottlenecks in modern two-stage retrieval systems. By formalizing the trade-off between retrieval effectiveness and system latency into a quantitative \textit{utility} function, we developed a pipeline that dynamically routes queries to the most cost-effective reranking strategy, ranging from skipping re-ranking entirely (BM25) to a heavy cross-encoder model (BGE). Our experiments demonstrated that the ``one-size-fits-all'' paradigm, where one reranker is applied always, is computationally inefficient. We showed that a significant portion of queries benefit from lightweight models without compromising retrieval quality. 
\vspace{-5px}
\section{Limitations \& Future Work}

\runinheading{Domain scope}
Due to computational and time constraints, our training dataset might not capture a sufficiently representative variety of queries and domains, therefore our learned router may not be fully general. Future work should curate a diverse dataset that focuses on clean, high-quality multidomain datasets that enable models to learn multiple domain-specific signals, if there are any. 

\runinheading{Model selection}
Our model selection was affected by the time and computational constraints. We also experimented with DistilBERT as a lighter router but observed a significant drop in accuracy, suggesting that our framework benefits from the full BERT capacity. Future work should explore other architectures and models.

\runinheading{Utility function}
We believe our utility function captures a good blend of effectiveness and latency of a selected model. Future work could refine the utility function, introduce new variables and more hyperparameters to adjust the weights according to personal needs. 
% although the utility easily be customized, future work could 

\runinheading{Results interpretation}
The router's test accuracy of $65\%$ leaves room for improvement. A diverse, balanced dataset might yield better results. Future iterations can implement a confidence thresholding mechanism in which low-confidence predictions can fall back to heavy reranker to ensure effectiveness.

%\runinheading{Acknowledgement}
%We thank members of UMass CIIR lab for their helpful feedback and support. 

% Our biggest limitation was our computational needs to be able to construct a sparser, more balanced dataset. We had to use smaller, datasets, sampling queries to keep it minimal feasible for our machines, causing us to have an imbalanced class distribution. Even though we saw high validation accuracies on our training setup with upsampling, this did not reflect on a real life scenario, as it induced a bias towards the minority, "heavier" class. Consequently, router selected the expensive and slower models for simpler queries, preventing us to achieve theoretical latencies shown. Additionally, our custom vocabulary and embeddings created from it heavily affected our ability to handle out-of-vocabulary terms effectively, opposed to pre-trained language models.
% \section{Future work}

% Future work will focus on explorting other architectures such as fine-tuned, lightweight Transformer models (e.g., DistilBERT) to leverage pre-trained knowledge for better generalization across diverse domains. We also plan to address the class imbalance issue more robustly by modifying our class-scoring functions, collecting more data points, rather than relying on simple upsampling. Finally, we intend to refine the utility function to include dynamic latency thresholds, allowing the system to adjust its routing sensitivity based on real-time hardware constraints or user-defined latency budgets.

\section*{Acknowledgements}
This work was supported in part by the Center for Intelligent Information Retrieval (CIIR). Any opinions, findings, conclusions, or recommendations expressed in this material are those of the authors and do not necessarily reflect those of the sponsor.

\balance
\bibliography{custom}
\bibliographystyle{ACM-Reference-Format}

\end{document}